\documentstyle[preprint,prd,eqsecnum,aps,epsf,axodraw]{revtex}

\tightenlines
\begin{document}

%============ The new commands: =============================
    \newcommand{\DSC}{D\hspace{-0.25cm}\slash_{\bot}}
    \newcommand{\DSP}{D\hspace{-0.25cm}\slash_{\|}}
    \newcommand{\DS}{D\hspace{-0.25cm}\slash}
    \newcommand{\DC}{D_{\bot}}
    \newcommand{\DSCX}{D\hspace{-0.20cm}\slash_{\bot}}
    \newcommand{\DSPX}{D\hspace{-0.20cm}\slash_{\|}}
    \newcommand{\DP}{D_{\|}}
    \newcommand{\QV}{Q_v^{+}}
    \newcommand{\QVB}{\bar{Q}_v^{+}}
    \newcommand{\QVP}{Q^{\prime +}_{v^{\prime}} }
    \newcommand{\QVBP}{\bar{Q}^{\prime +}_{v^{\prime}} }
    \newcommand{\QVHZ}{\hat{Q}^{+}_v}
    \newcommand{\QVHZB}{\bar{\hat{Q}}_v{\vspace{-0.3cm}\hspace{-0.2cm}{^{+}} } }
    \newcommand{\QVPHZB}{\bar{\hat{Q}}_{v^{\prime}}{\vspace{-0.3cm}\hspace{-0.2cm}{^{\prime +}}} }
    \newcommand{\QVPHFB}{\bar{\hat{Q}}_{v^{\prime}}{\vspace{-0.3cm}\hspace{-0.2cm}{^{\prime -}} } }
    \newcommand{\QVPHB}{\bar{\hat{Q}}_{v^{\prime}}{\vspace{-0.3cm}\hspace{-0.2cm}{^{\prime}} }   }
    \newcommand{\QVHF}{\hat{Q}^{-}_v}
    \newcommand{\QVHFB}{\bar{\hat{Q}}_v{\vspace{-0.3cm}\hspace{-0.2cm}{^{-}} }}
    \newcommand{\QVH}{\hat{Q}_v}
    \newcommand{\QVHB}{\bar{\hat{Q}}_v}
    \newcommand{\VS}{v\hspace{-0.2cm}\slash}
    \newcommand{\MQ}{m_{Q}}
    \newcommand{\MQP}{m_{Q^{\prime}}}
    \newcommand{\QVHPMB}{\bar{\hat{Q}}_v{\vspace{-0.3cm}\hspace{-0.2cm}{^{\pm}} }}
    \newcommand{\QVHMPB}{\bar{\hat{Q}}_v{\vspace{-0.3cm}\hspace{-0.2cm}{^{\mp}} }  }
    \newcommand{\QVHPM}{\hat{Q}^{\pm}_v}
    \newcommand{\QVHMP}{\hat{Q}^{\mp}_v}
 %==============================================================

\draft
%\preprint{AS-ITP-99-06, hep-ph/ }
%\preprint{AS-ITP-?-?, hep-ph/? }
\title{ $B\to\pi l \nu$ Decay and $|V_{ub}|$}
\author{ W.Y. Wang and Y.L. Wu }
\address{Institute of Theoretical Physics, Chinese Academy of Sciences\\ Beijing 100080, China }
%\date{}
\maketitle

\begin{abstract}
$B\to\pi l \nu$ decay is studied in the effective theory of heavy
quark with infinite mass limit. The leading order heavy
flavor-spin independent universal wave functions which
parametrize the relevant matrix elements are evaluated via light
cone sum rule method in the effective theory. The important quark
mixing matrix element $|V_{ub}|$ is then extracted via $B\to\pi
l\nu$ decay mode.
\end{abstract}

\pacs{PACS numbers: 13.20.-v, 11.30.Hv, 11.55.Hx\\
 Keywords: heavy
to light transition, $|V_{ub}|$, effective
 theory, light cone sum rule}

\newpage

\section{Introduction}\label{int}

Weak decays of charmed and beautiful
hadrons are quite favorable in particle physics because of their usage in
determining fundamental parameters of the standard model and testing various
theories and models.
Among these heavy hadron decays the semileptonic decays $B\to \pi l\nu$ and $B\to\rho l\nu$
have been observed experimentally. These exclusive decays provide one of the main channels to
determine the important CKM matrix element $|V_{ub}|$.

The difficulty in studying $B\to\pi l\nu$ and $B\to\rho l\nu$
decays mainly concerns the calculation of the relevant hadronic
matrix elements of weak operators, or, equivalently, the
corresponding form factors which contain nonperturbative
contributions as well as perturbative ones and are beyond the power of
pure QCD perturbation theory. Up to present these form factors are
usually evaluated from lattice calculations, QCD sum rules and
some hadronic models.

Sum rule method has been applied to $B\to\pi(\rho) l\nu$ decay in
the full QCD and provided reasonable results\cite{var,ar,arsc}.
Since the meson B contains a single heavy quark, it is expected
that its exclusive decays into light mesons may also be understood
well in the effective theory of heavy quark, which explicitly
demonstrates the heavy quark spin-flavor symmetry and its breaking
effects can systematically be evaluated via the power of inverse
heavy quark mass $1/m_Q$. The effective theory of heavy quark has
been widely applied to heavy hadron systems, such as B decays into
heavy hadrons via both exclusive and inclusive decay modes. There
are two different versions of effective theory of heavy quark. One
is the heavy quark effective theory (HQET), which generally
decouples the "quark fields" and "antiquark fields" and treats one
of them independently. This treatment is only valid when taking
the heavy quark mass to be infinite. In the real world, mass of
quark must be finite, thus one should keep in the effective
Lagrangian both the effective quark and effective antiquark
fields. Based on this consideration, a heavy quark effective field
theory (HQEFT) \cite{ylw,wwy,yww,wy,ww,excit} has been established
and investigated with including the effects of the mixing terms
between quark and antiquark fields. Its applications to the pair
annihilation and creation have also been studied in the
literature\cite{spain1,spain2,spain3}. Though the HQEFT explicitly
deviate from HQET from the next-to-leading order, these two
formulations of effective theory trivially coincide with each
other at the infinite heavy quark mass limit. In our knowledge the
exclusive heavy to light (pseudoscalar) decay channels have been
discussed in \cite{gzmy}, where  the matrix elements in the
effective theory have been formulated, but the two leading order
wave functions have not been calculated.

In this paper we focus on the calculation of the leading
order wave functions of $B\to\pi l\nu$ decay by using the light
cone sum rule in the effective theory of heavy quark. As an
important application, $|V_{ub}|$ is extracted. In section 2, the
heavy to light matrix element is represented by two heavy quark
independent wave functions A and B. In section 3, we derive the
light cone sum rules for the calculation of A and B. In section
4, we present the numerical results and extract $|V_{ub}|$. Our
short summary is drawed in the last section.

 \section{B $\to \pi l\nu$ decay matrix element}\label{formulation}

The matrix elements responsible for $B\to\pi l\nu$ decay is
$<\pi(p)|\bar{u}\gamma^\mu b|B>$, where b is the beautiful quark field in full QCD.
It is generally parametrized by two form factors as follows,
\begin{eqnarray}
\label{fdef}
  <\pi(p)|\bar u\gamma^\mu b|B(p+q)>=2f_{+}(q^2) p^\mu+(f_{+}(q^2)+f_{-}(q^2))q^\mu.
\end{eqnarray}

In the effective theory of heavy quark, matrix elements can be analyzed order
by order in powers of
the inverse of the heavy quark mass $1/m_Q$ and also be conveniently
expressed by some heavy spin-flavor indenpendent universal wave functions
\cite{wwy,ww,excit,gzmy}.

Here we adopt the following normalization of the matrix elements in full QCD
and in the effective theory \cite{wwy,ww,excit}:
\begin{eqnarray}
\label{normalization}
\frac{1}{\sqrt{m_B} }<\pi(p)|\bar u \Gamma b|B>=\frac{1}{\sqrt{\bar\Lambda_B}}
\{  <\pi(p)|\bar{u} \Gamma \QV|B_v>+O(1/m_b) \},
\end{eqnarray}
where $\bar\Lambda_B=m_B-m_b$,
and $$\bar\Lambda=\lim_{m_Q\to \infty} \bar\Lambda_B$$ is the heavy
flavor independent binding energy reflecting the effects of the light degrees
of freedom in the heavy hadron. $\QV$ is the effective heavy quark field in effective
theory.

Associate the heavy meson state with the spin wave function
\begin{eqnarray}
 \label{eq:spinwave1}
 {\cal M}_v=\sqrt{\bar{\Lambda}} \frac{1+v\hspace{-0.2cm}\slash}{2} \left\{
\begin{array}{ll}
 -\gamma_{5} & \mbox{for pseudoscalar meson} \\
 \epsilon\hspace{-0.15cm}\slash & \mbox{for vector meson with polarization vector}\; \; \epsilon^\mu
\end{array}
 \right.
\end{eqnarray}
we can analyze the matrix element in effective theory by carrying
out the trace formula :
\begin{eqnarray}
\label{parinhqet}
  <\pi(p)|\bar{u} \Gamma \QV|B_v>=-Tr[\pi(v,p)\Gamma {\cal M}_v]
\end{eqnarray}
with
\begin{eqnarray}
\label{ABdef}
 \pi(v,p)&=&\gamma^5 [A(v\cdot p,\mu)+ {\hat{p}\hspace{-0.2cm}\slash}
B(v\cdot p,\mu)], \nonumber\\  \hat{p}^\mu&=&\frac{p^\mu}{v\cdot p} \;\;.
\end{eqnarray}
A and B are the leading order wave functions characterizing the
heavy-to-light-pseudoscalar transition matrix elements in the
effective theory. They are heavy quark mass independent, but are
functions of the variable $v\cdot p$ and the energy scale $\mu$ as
well. Nevertheless, since the discussion in the present paper is
rrestricted within the tree level, we neglect the $\mu$ dependence
from now on.

Combining eqs. (\ref{fdef})-(\ref{ABdef}), one gets
\begin{eqnarray}
\label{relation}
 f_{\pm}(q^2)&=&\frac{1}{\sqrt{m_b}} \sqrt{ \frac{m_B \bar\Lambda}
 {m_b \bar{\Lambda}_B } }
    \{ A(v\cdot p)\pm B(v\cdot p) \frac{m_b}{v\cdot p}  \} +\cdots,
\end{eqnarray}
%\begin{eqnarray}
%\label{relation}
% f_{+}(q^2)&=&\sqrt{ \frac{m_B \bar\Lambda}{\bar\Lambda_B}  }
%     (A(v\cdot p)/m_b+B(v\cdot p)/(v\cdot p))+\cdots ;\nonumber\\
% f_{-}(q^2)&=&\sqrt{ \frac{m_B \bar\Lambda}{\bar\Lambda_B}  }
%     (A(v\cdot p)/m_b-B(v\cdot p)/(v\cdot p))+\cdots ,
%\end{eqnarray}
where the dots denote higher order $1/m_Q$ contributions which
will not be taken into account in the present paper. Note that we
have used different variables for $f_+$, $f_-$ and $A$, $B$. The
relation between the variables $v\cdot p$ and $q^2$ is
\begin{equation}
\label{ydef}
y\equiv v\cdot p=\frac{m^2_B+m^2_\pi-q^2}{2m_B}.
\end{equation}

 \section{Light cone sum rule for $B\to\pi l\nu$}\label{sumrule}

The QCD sum rule based on short distance expansion has been proved
to be quite fruitful in solving a variety of hadron problems.
Nevertheless, it is also well known that this method meets
difficulties in the case of heavy to light transition because the
coefficients of the subleading quark and quark-gluon condensate
with the heavy quark mass terms grow faster than the perturbative
contribution, which implies the breakdown of the short distance
operator product expansion (OPE) in the heavy mass limit.
Alternatively, it has been found that heavy to light decays can be
well studied by light cone sum rule approach, in which the
corresponding correlators are expanded near the light cone in
terms of meson wave functions. In this way the nonperturbative
contributions are embeded in the meson wave functions instead of
the vacuum condensates in the short distance OPE sum rule. Though
there are some differences in the techniques of calculation, the
two sum rule methods are based on the same idea of quark-hadron
duality and dispersion relation, and furthermore, they follow the
same procedure in deriving form factors.

For $B\to\pi l\nu$ decay, one may consider the vacuum-pion correlation function
\begin{eqnarray}
\label{correlator}
F^\mu(p,q)=i\int d^4x e^{iq\cdot x} <\pi(p)|T\{\bar{u}(x)\gamma^\mu b(x),\bar{b}(0)
  i\gamma^5 d(0)\} |0>.
\end{eqnarray}
Here $p$ and $q$ are momenta carried by the pion and leptons. The B meson has
momentum $P_B=p+q$.
Inserting a complete set of states with B meson quantum numbers, we obtain the phenomenological
representation
\begin{eqnarray}
\label{cor}
F^\mu(p,q)_{phen}=\frac{<\pi(p)|\bar{u}\gamma^\mu b|B><B|\bar{b}i\gamma^5 d|0>}
   { m^2_B-(p+q)^2 }+\sum_H \frac{<\pi(p)|\bar{u}\gamma^\mu b|H><H|\bar{b}i\gamma^5 d|0>}
   { m^2_H-(p+q)^2 }.
\end{eqnarray}
With the normalization relation in (\ref{normalization}), the matrix elements in
(\ref{cor}) can be expanded into  the ones in effective theory of heavy quark in
powers of $1/m_b$. When all higher $1/m_b$ order contributions are neglected,
(\ref{cor}) reduces straightforwardly into
\begin{eqnarray}
\label{phen}
  2 i F \frac{ A v^\mu+B \hat{p}^\mu }{2\bar\Lambda_B-2v\cdot k}
    +\int^{\infty}_{s_0} ds \frac{\rho(v\cdot p,s)}{s-2v\cdot k}+Subtractions,
\end{eqnarray}
where $k^\mu$ is the heavy hadron's residual momentum, $k^\mu=P^\mu_B-m_b v^\mu$.
The first term in (\ref{phen}) is a pole contribution obtained by using
(\ref{parinhqet}) together with the parametrization
\begin{eqnarray}
\label{Fdef}
<0|\bar{q} \Gamma \QV |B_v >=\frac{F}{2}Tr[\Gamma {\cal M}_v]
\end{eqnarray}
with F being the leading order decay constant of B meson in effective theory \cite{ww}.
The second term in (\ref{phen}) is the higher resonance contributions given in the form
of an integral over the physical spectral density $\rho(v\cdot p,s)$.
Note that the Lorentz indices of the second term in (\ref{phen}) are not written
explicitly but embeded in $\rho(v\cdot p,s)$.

On the other hand, the correlator can be calculated and expressed as the form of an integration
over the theoretic spectral density $\rho(v\cdot p,s)_{theory}$, which equals
to $\rho(v\cdot p,s)$ under the assumption of quark-hadron duality. Namely,
the correlator (\ref{correlator}) can be written as
\begin{eqnarray}
\label{theo}
\int^{\infty}_{0} ds \frac{\rho(v\cdot p,s)}{s-2v\cdot k}+Subtractions.
\end{eqnarray}
Equating (\ref{phen}) and (\ref{theo}) yields
\begin{eqnarray}
\label{phentheo}
  2iF \frac{ A v^\mu+B \hat{p}^\mu }{2\bar\Lambda_B-2v\cdot k}
   =\int^{s_0}_{0} ds \frac{\rho(v\cdot p,s)}{s-2v\cdot k}+Subtractions.
\end{eqnarray}

So the next step involves the calculation of (\ref{correlator}) in the
framework of effective theory of heavy quark.
Substituting the heavy hadron states and heavy quark fields into
the effective ones in the effective theory, and then performing the corresponding momentum
shift $P^\mu_B-m_b v^\mu=k^\mu$, we obtain when neglecting higher $1/m_Q$ order
corrections
\begin{eqnarray}
\label{correlatorinHQET}
  F^\mu(p,q)=i\int d^4x e^{i(q-m_bv)\cdot x}
  <\pi(p)|T{\bar{u}(x)\gamma^\mu \QV(x),      \QVB(0)i\gamma^5 d(0) }|0>.
\end{eqnarray}
In the light cone sum rule approach, one should contract the heavy quark fields and expand the
correlator into a series in powers of the twist of light cone pion wave functions.
These light cone wave functions provide an alternative treatment besides the vacuum
condensates. They have been discussed in detail in many references \cite{var,ar,vvar}.
Up to twist 4, the pion wave functions relevant to $B\to\pi l\nu$ decay are defined as follows,
\begin{eqnarray}
<\pi(p)|\bar{u}(x)\gamma^\mu \gamma^5 d(0)|0>&=&-ip^\mu f_\pi \int^1_0 du e^{iup\cdot x}
   [\phi_\pi(u)+x^2 g_1(u) ]\nonumber\\
   &+&f_\pi (x^\mu-\frac{x^2 p^\mu}{x\cdot p})
   \int^1_0 du e^{iup\cdot x} g_2(u), \nonumber\\
<\pi(p)|\bar{u}(x)i \gamma^5 d(0)|0>&=&\frac{f_\pi m^2_\pi}{m_u+m_d} \int^1_0 du
   e^{iup\cdot x} \phi_p(u), \nonumber\\
<\pi(p)|\bar{u}(x) \sigma_{\mu\nu} \gamma^5 d(0)|0>&=& i(p_\mu x_\nu-p_\nu x_\mu)
   \frac{f_\pi m^2_\pi}{6 (m_u+m_d) } \int^1_0 du e^{iup\cdot x} \phi_\sigma(u).
\end{eqnarray}
$\phi_\pi$ is the leading twist 2 wave function. $\phi_p$ and $\phi_\sigma$ are twist 3
wave functions, while $g_1$ and $g_2$ are wave functions of twist 4.

Using the propagator $\frac{1+v\hspace{-0.13cm}\slash}{2} \int^{\infty}_{0} dt \delta(x-vt) $
for the contraction of the effective heavy quark fields, we get
\begin{eqnarray}
\label{corresult}
F^\mu(y,\omega)&=&-\frac{i f_{\pi}}{2} \int^{\infty}_{0} dt \int^{1}_{0} du e^{\frac{it\omega}{2}}
  e^{i y t (u-1)} \{  v^\mu [tg_2(u)-i\mu_\pi \phi_p(u)-\frac{t}{6}\mu_\pi y \phi_\sigma(u) ]
  \nonumber\\
  &+&\hat{p}^\mu y[-i\phi_\pi-it^2 g_1(u)-\frac{t}{y}g_2(u)+\frac{t}{6}\mu_\pi \phi_\sigma(u) ]\}
\end{eqnarray}
with $\omega \equiv 2v\cdot k$ and $\mu_\pi\equiv \frac{m_\pi^2}{(m_u+m_d)}$.

In order to proceed, we perform a wick rotation of the t axis and then apply the Borel
transformation $\hat{B}^{(\omega)}_T$ to (\ref{corresult}). The result is
\begin{eqnarray}
\label{BTcor}
\hat{B}^{(\omega)}_T F^\mu(y,\omega)=-i f_\pi \int^1_0 du e^{\frac{2y}{T}(u-1)} \{ v^\mu
   [-\frac{2}{T}g_2(u)-\mu_\pi \phi_p(u)+\frac{1}{3T}\mu_\pi y\phi_\sigma(u)] \nonumber\\
   +\hat{p}^\mu y[-\phi_\pi(u)+\frac{4}{T^2}g_1(u)+\frac{2}{yT}g_2(u)-\frac{1}{3T}\mu_\pi
   \phi_\sigma(u) ]\}.
\end{eqnarray}
In deriving this equation we have used the feature of Borel transformation:
\begin{eqnarray}
   \hat{B}^{(\omega)}_T e^{\lambda \omega}=\delta(\lambda-\frac{1}{T}).
\end{eqnarray}

It is found that the spectral function used in sum rule can be obtained by performing
a continuous double
Borel transformation on the amplitude itself \cite{nr,pv}. In order to get the spectral
function $\rho(y,s)$, we now carry out the continuous Borel transformations as follows
\begin{eqnarray}
  \rho(y,s)=\hat{B}^{(-1/T)}_{1/s} \hat{B}^{(\omega)}_T F^\mu(y,\omega).
\end{eqnarray}
The result is
\begin{eqnarray}
\label{spectralfun}
 \rho(y,s)&=&-\frac{i f_\pi}{2y} \{v^\mu [\frac{1}{y} \frac{\partial}{\partial u}g_2(u)
    -\mu_\pi \phi_p(u)-\frac{\mu_\pi}{6}\frac{\partial}{\partial u}\phi_\sigma(u) ]
    +\hat{p}^\mu y[-\phi_{\pi} (u)+\frac{1}{y^2}\frac{\partial^2}{\partial u^2}g_1(u) \nonumber\\
    &-&\frac{1}{y^2} \frac{\partial}{\partial u} g_2(u)
    +\frac{\mu_\pi}{6y}\frac{\partial}{\partial u} \phi_\sigma(u) ]\}_{u=1-\frac{s}{2y}}.
\end{eqnarray}
In the derivation of (\ref{spectralfun}), $\frac{1}{T}$ has been first expressed
as a derivative of the
exponent in (\ref{BTcor}) over u, and then the method of integration by parts
over u has been used.

(\ref{phentheo}) and (\ref{spectralfun}) immediately yield:
\begin{eqnarray}
\label{sr}
 A(y)&=&-\frac{f_\pi}{4 F} \int^{s_0}_{0} ds e^{\frac{ 2\bar\Lambda_B-s}{T}}
    [\frac{1}{y^2} \frac{\partial}{\partial u}g_2(u)-\frac{\mu_\pi}{y} \phi_p(u)
     -\frac{\mu_\pi}{6y}
    \frac{\partial}{\partial u}\phi_\sigma(u) ]_{u=1-\frac{s}{2y}},\nonumber\\
 B(y)&=&-\frac{f_\pi}{4 F} \int^{s_0}_{0} ds e^{\frac{ 2\bar\Lambda_B-s}{T}}
    [-\phi_{\pi} (u)+\frac{1}{y^2}\frac{\partial^2}{\partial u^2}g_1(u)
    -\frac{1}{y^2} \frac{\partial}{\partial u} g_2(u)
    +\frac{\mu_\pi}{6y}\frac{\partial}{\partial u} \phi_\sigma(u) ]_{u=1-\frac{s}{2y}},
\end{eqnarray}

\section{Numerical Results}\label{sum}

For the light cone wave functions appearing in the sum rules (\ref{sr}),
we take \cite{ar,vvar,vi}
\begin{eqnarray}
\label{pifunction}
\phi_\pi(u)&=& 6u(1-u)\{1+\frac{3}{2}a_2[5(2u-1)^2-1]+\frac{15}{8}a_4
  [21 (2u-1)^4-14 (2u-1)^2+1]\} , \nonumber\\
\phi_p(u)&=& 1+\frac{1}{2}B_2 [3(2u-1)^2-1]+\frac{1}{8}B_4 [35 (2u-1)^4
  -30 (2u-1)^2+3] ,\nonumber\\
\phi_\sigma(u)&=& 6u(1-u)\{ 1+\frac{3}{2}C_2 [5(2u-1)^2-1]+\frac{15}{8}C_4
  [21 (2u-1)^4-14(2u-1)^2+1]\},  \nonumber\\
g_1(u)&=& \frac{5}{2}\delta^2 u^2 (1-u)^2+\frac{1}{2}\epsilon \delta^2
  [u(1-u)(2+13u (1-u)+10u^3 \log u(2-3u+\frac{6}{5}u^2) \nonumber\\
  &+&10(1-u)^3 \log((1-u)(2-3(1-u)+\frac{6}{5}(1-u)^2)) ], \nonumber\\
g_2(u)&=&\frac{10}{3} \delta^2 u(1-u)(2u-1).
\end{eqnarray}
The asymptotic form of these functions and the scale dependence are given by
perturbative QCD \cite{va,bf}.

For the convenience of comparison, we use the same values for the
parameters as in \cite{ar,vvar},
\begin{eqnarray}
\label{para}
a_2(\mu_b)=0.35, \;\; a_4(\mu_b)=0.18, \;\; B_2(\mu_b)=0.29, \;\;
B_4(\mu_b)=0.58,   \nonumber\\
C_2(\mu_b)=0.059, \;\; C_4(\mu_b)=0.034, \;\;
\delta^2(\mu_b)=0.17 \mbox{GeV}^2, \;\; \epsilon(\mu_b)=0.36 .
\end{eqnarray}
$\mu_b$ is the appropriate scale set by the typical virtuality of the beautiful
quark,
\begin{equation}
\label{mub}
\sqrt{m^2_B-m^2_b} \approx 2.4\mbox{GeV}.
\end{equation}

Besides all these parameters the numerical analysis of the sum rules
(\ref{sr}) needs also the hadron quantities $\mu_\pi$, $f_\pi$,
$\bar\Lambda_B$ and $F$. These quantities have been studied via sum rules
and other approaches by several groups.
With the values $\mu_\pi=2.02$GeV, $f_\pi=0.132$GeV\cite{ar,vvar},
$\bar\Lambda_B=0.53$GeV and $F=0.30\mbox{GeV}^{3/2}$ \cite{ww}, we get from
eqs.(\ref{sr}) the results for A and B given in the figures Fig.1-4. In these
figures A and B are shown as functions of T and $y=v\cdot p$.
We are mainly interested in the range of $T=2.0\pm 1.0 $GeV, where
both the twist 4 corrections and the contributions from excited and continuum
states do not exceed 30\%.
It is seen that the curves in Fig.1 and
Fig.2 are quite stable in this range for the threshold energy $s_0=2.3\pm 0.6$GeV.
In Fig.3 and Fig.4, A and B
become rather stable with respect to the variation of $y=v\cdot p$ when $y>1.5$GeV.
However, they become unstable at small $y$, which corresponds to large momentum
transfer $q^2$. This is in expectation because the light cone expansion and the
sum rule method would break down as $q^2$ approaches near $m^2_b$ \cite{ar}.

We also derive $f_+(q^2)$ and $f_-(q^2)$ from $A(v\cdot p)$ and $B(v\cdot p)$
by using the relations in (\ref{relation}) and the beautiful quark mass
$m_b=m_B-\bar\Lambda_B=4.75$GeV. The results are shown in Fig.5-6. It is readly seen
that when the momentum transfer $q^2$ grows large (e.g. over $16\mbox{GeV}^2$ for
the curve of $s_0=2.3$GeV in Fig.5-6), the values
of $f_+$ and $f_-$ derived from sum rules become rather unstable and should not be
trusted.

In order to predict the decay width and $|V_{ub}|$, one should have knowledge
on the behavior of form factors in the whole kinematically accessible region.
Now for large momentum transfer we have the single pole approximation \cite{ar}
\begin{eqnarray}
\label{sinpole}
f_+(q^2)=\frac{f_{B^*} g_{B^*B\pi}}{2m_{B^*}(1-q^2/m^2_{B^*})  }.
\end{eqnarray}
The couplings $f_{B^*}$ and $g_{B^*B\pi}$ have been studied in previous
papers. Here we would use $m_{B^*}=5.325$GeV, $f_{B^*}=0.16\pm 0.03$GeV
and $g_{B^*B\pi}=29\pm 3$\cite{ar}.

Next we write $f_+(q^2)$ as
\begin{eqnarray}
\label{fitform}
f_+(q^2)=\frac{f_+(0)}{1-a q^2/m^2_B+b q^4/m^4_B}
\end{eqnarray}
and fit the parameters $a$ and $b$ by using the sum rules and eq. (\ref{sinpole}).
For the threshold $s_0=2.3$GeV, we
choose proper $a$ and $b$ to make (\ref{fitform}) approach the
sum rule results at $q^2<15\mbox{GeV}^2$ but compatible with eq. (\ref{sinpole}) at
$q^2>15\mbox{GeV}^2$. Our favorable parameters are
\begin{eqnarray}
\label{ab}
a=1.31, \;\; b=0.35, \;\; f_+(0)=0.35.
\end{eqnarray}

The values of $f_+(q^2)$ at $T=2.0$GeV calculated from
(\ref{sinpole}), (\ref{fitform}) and the light cone sum rules are
shown in Fig.7. It is found that the single pole model
extrapolation matches quite well with the direct estimation from
our light cone sum rules (\ref{sr}) at intermediate momentum
transfer around $q^2=15\mbox{GeV}^2$. This implies that our
discription of $f_+(q^2)$ by (\ref{sinpole}) together with the sum
rules (\ref{sr}) (but in different applicable regions) is
self-consistent.

For the lepton $l=e \;\;\mbox{or}\;\; \mu$, the lepton mass $m_l$
may be safely neglected, and the decay width of $B\to \pi l\nu$
has the distribution on momentum transfer $q^2$ as follows
 \begin{eqnarray}
\label{gammaq2}
\frac{d\Gamma}{dq^2}=\frac{G^2_F |V_{ub}|^2}{24 \pi^3} (E^2_\pi-m^2_\pi)^{3/2}
   [f_+(q^2)]^2.
\end{eqnarray}
Here $E_\pi=y=(m^2_B+m^2_\pi-q^2)/(2m_B)$ is the pion energy in the B meson rest frame.

With the pion mass $m_\pi=0.14$GeV and the parametrizations of (\ref{fitform}),
we obtain the integrated width
\begin{eqnarray}
\label{width}
\Gamma(B\to \pi l \nu)=(10.2\pm 1.5) |V_{ub}|^2 \mbox{ps}^{-1}.
\end{eqnarray}
The error in eq.(\ref{width}) results from the variation of the threshold energy
in $s_0=1.7-2.9$GeV.

From the branching fraction measured by CLEO collaboration \cite{cleo},
$\mbox{Br}(B^0 \to \pi^- l^+ \nu_l)=(1.8\pm 0.4\pm 0.3\pm 0.2) \times 10^{-4}$
and the world average of the $\mbox{B}^0$ lifetime \cite{pdg},
$\tau_{\tiny{\mbox{B}^0}}=1.56\pm 0.06 \;\mbox{ps}$, one has \cite{ar}
\begin{eqnarray}
\label{CLEO}
\Gamma(B^0\to \pi^- l^+ \nu_l)=(1.15\pm 0.35)\times 10^{-4} \mbox{ps}^{-1}.
\end{eqnarray}
Comparison of (\ref{width}) and (\ref{CLEO}) yields
\begin{eqnarray}
|V_{ub}|=(3.4\pm 0.5 \pm 0.3)\times 10^{-3},
\end{eqnarray}
where the first (second) error corresponds to the experimental (theoretical)
 uncertainty. Here the theoretical uncertainty is mainly
 considered from the threshold effects. In general, higher order
 contributions need to be included for all the relevant
 parameters.
It was noticed that the two-loop QCD perturbative correction may
be significant for an accurate determination of B meson decay
constants \cite{ww}. In particular, it may enlarge the constant F
by about 25\%, and increase $\bar\Lambda$ at the same time. These
effects evidently worsen the accuracy of our extraction of
$|V_{ub}|$. By taking into account this uncertainty, we arrive at
the following result
\begin{eqnarray}
\label{vub}
|V_{ub}|=(3.4\pm 0.5 \pm 0.5)\times 10^{-3},
\end{eqnarray}

This estimate is in good agreement with that derived from full QCD
calculation \cite{ar}:
\begin{eqnarray}
\label{vubQCD}
|V_{ub}|&=&(3.9\pm0.6\pm 0.6 )\times 10^{-3} \;\; (\mbox{via}\;\; B\to\pi l \nu),\nonumber\\
|V_{ub}|&=&(3.4\pm0.6\pm 0.5 )\times 10^{-3} \;\; (\mbox{via}\;\; B\to\rho l \nu).
\end{eqnarray}
Furthermore, the value of $|V_{ub}|$ obtained in eq.(\ref{vub}) is
also close to the one given by CLEO \cite{cleo2},
\begin{equation}
\label{CLEOcomb}
|V_{ub}|=(3.25\pm 0.14^{+0.21}_{-0.29} \pm 0.55)\times 10^{-3},
\end{equation}
which is a combined result from the analyses based on different models and treatments
on $B\to \pi (\rho) l \nu$ transitions.

\section{Summary}\label{end}

In this paper we have studied $B\to \pi l\nu$ decay by using the
light cone sum rule approach within the framework of effective
theory for heavy quark. Two leading order wave functions in the
effective theory with infinite mass limit have been calculated.
The important CKM matrix element $|V_{ub}|$ has been extracted and
its value has been found to be
\begin{eqnarray}
|V_{ub}|=(3.4\pm 0.5 \pm 0.5)\times 10^{-3}.
\end{eqnarray}
It has been seen that the value of $|V_{ub}|$ extracted from the
leading order heavy quark expansion coincides well with that
extracted from the full QCD calculation, which shows the
reliability of the heavy quark expansion and the power of light
cone sum rule approach in studying heavy to light exclusive
decays. Working out $1/m_Q$ contributions should be interesting,
and it is expected to cast more light on the treatment of heavy to
light decays by applying for the effective theory of heavy quark.

\acknowledgments

This work was supported in part by the NSF of China under the
grant No. 19625514  as well as Chinese Academy of Sciences.

%\begin{thebibliography}{99}

%\end{thebibliography}

\newpage
\centerline{\large{FIGURES}}
%================================

\newcommand{\PIC}[2]
{
\begin{center}
\begin{picture}(300,270)(0,0)
%%\put{
%\put(40,25){
%\epsfxsize=8cm
%\epsfysize=8cm

\put(0,5){
\epsfxsize=10cm
\epsfysize=10cm
\epsffile{#1} }
%\put(200,0){\makebox{#2}}
%\put(150,40){\makebox(0,0){#2}}
\put(150,90){\makebox(0,0){#2}}
\end{picture}
\end{center}
}

%\newpage
%\large{\centerline{FIGURES}}
\small
\mbox{}
{\vspace{1.2cm}}

\PIC{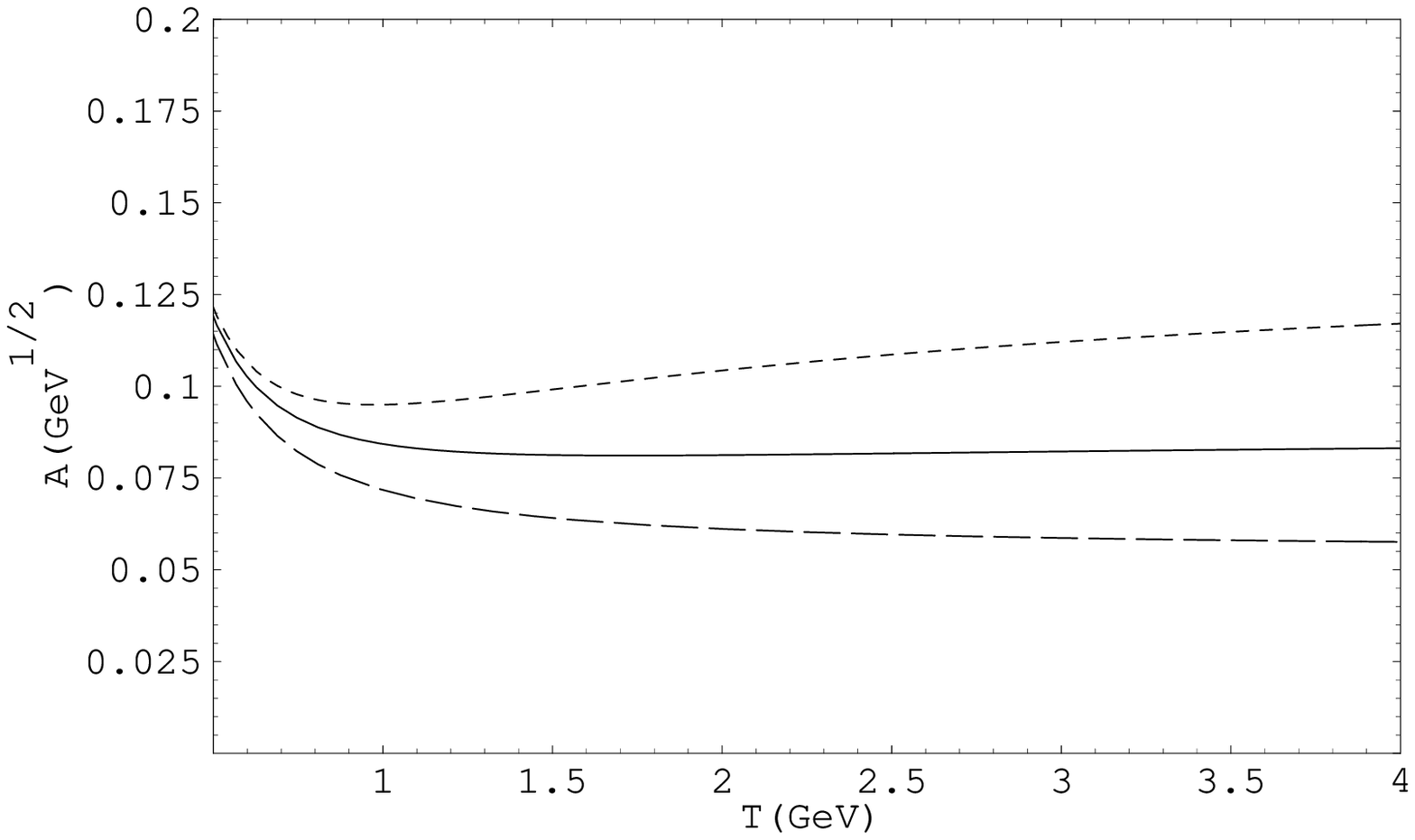}{Fig.1}

\PIC{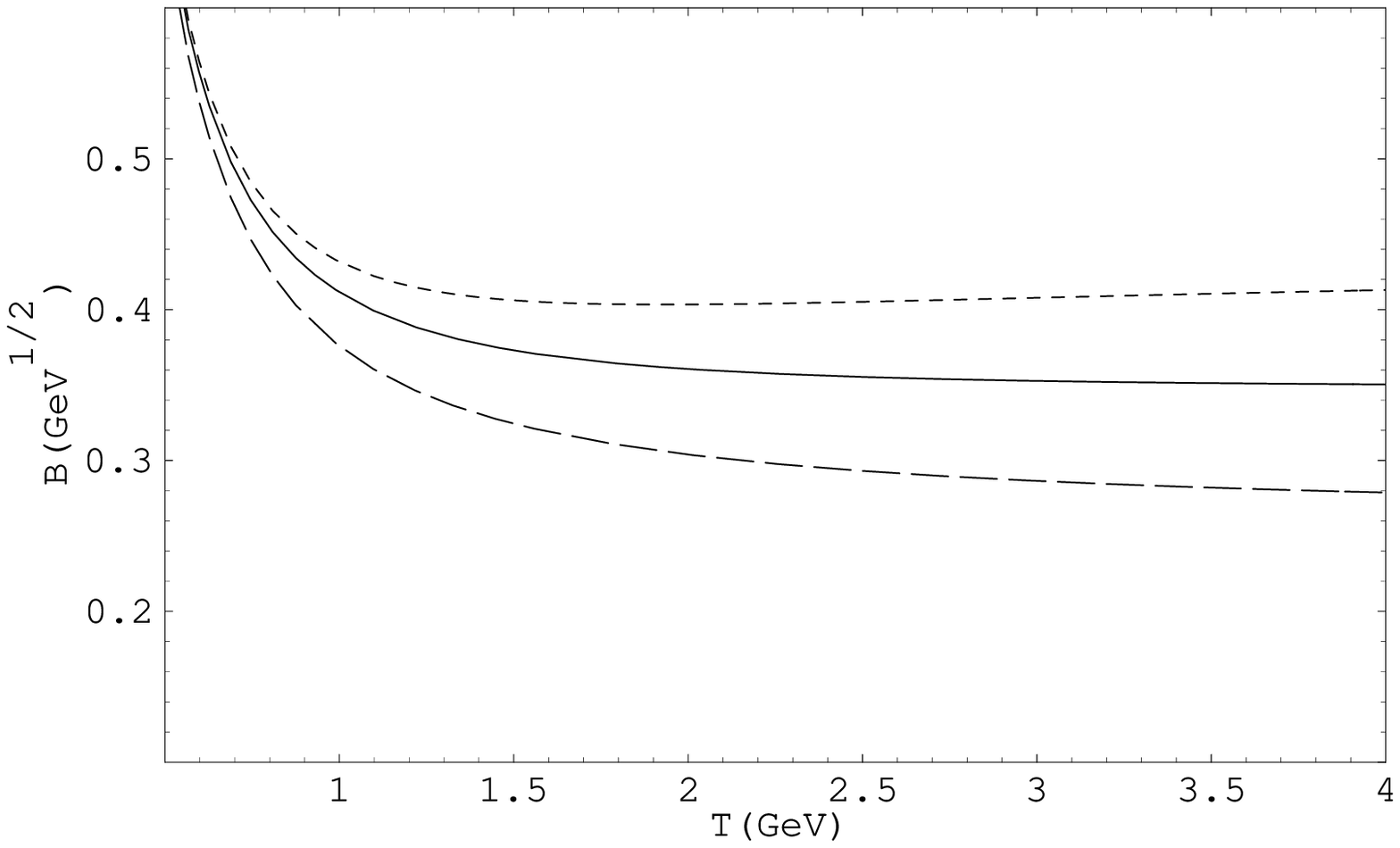}{Fig.2}

\vspace{-2cm}{\center{Fig.1-2. Variation of A and B with the Borel parameter T
for different values of the continuum threshold $s_0$. The dashed, solid and
dotted curves correspond to $s_0=$1.7, 2.3 and 2.9 GeV respectively.
$y=v\cdot p=2.64$ GeV is fixed, which corresponds to $q^2=0 \mbox{GeV}^2$.}}

\newpage
\mbox{}
\vspace{2cm}

\PIC{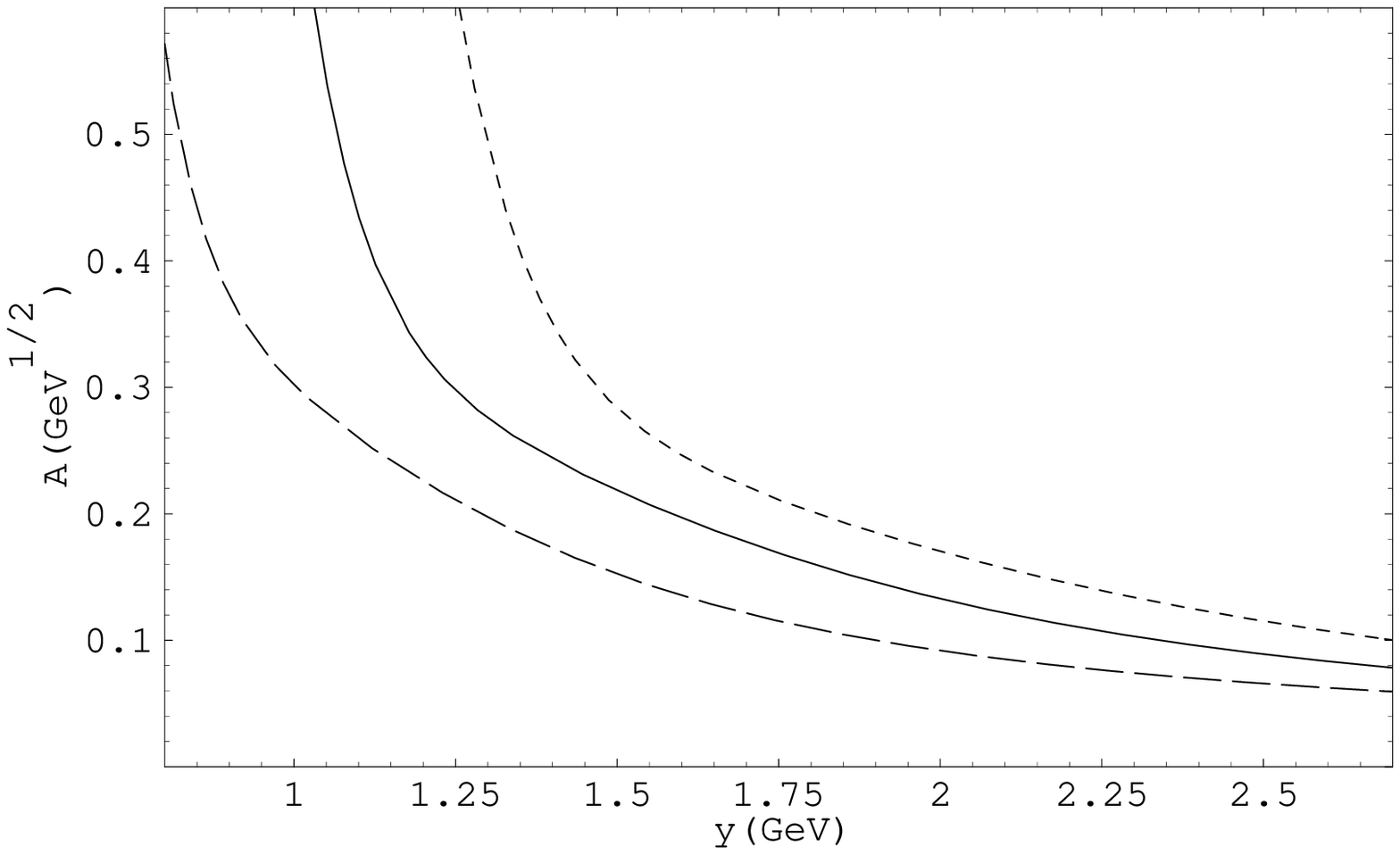}{Fig.3}

\PIC{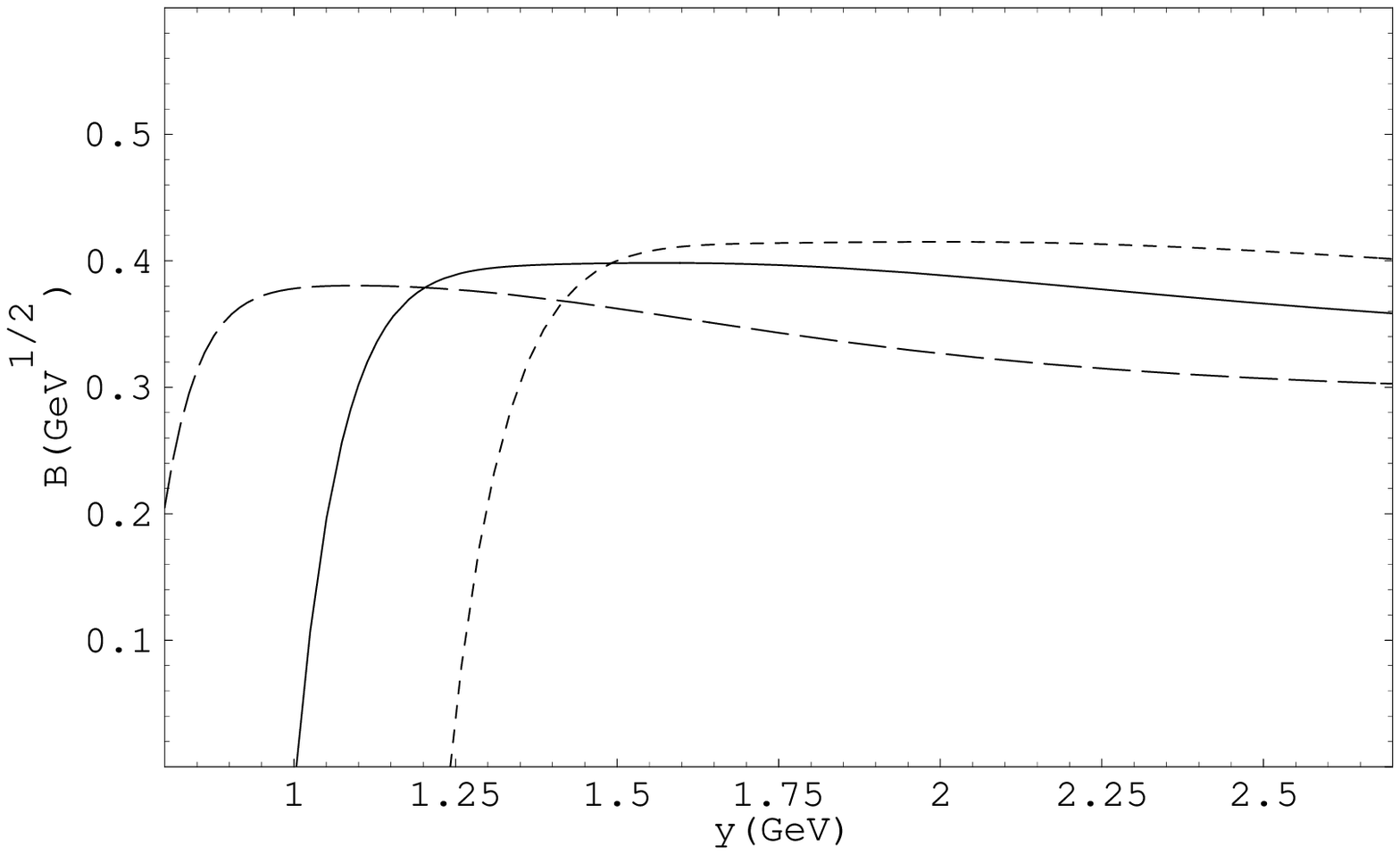}{Fig.4}

\vspace{-2cm}{\center{Fig.3-4. A and B as functions of $y=v\cdot p$
for different values of the continuum threshold $s_0$. The dashed, solid and
dotted curves correspond to $s_0=$1.7, 2.3 and 2.9 GeV respectively.
T=2.0 GeV is fixed. } }

\newpage
\mbox{}
\vspace{2cm}

\PIC{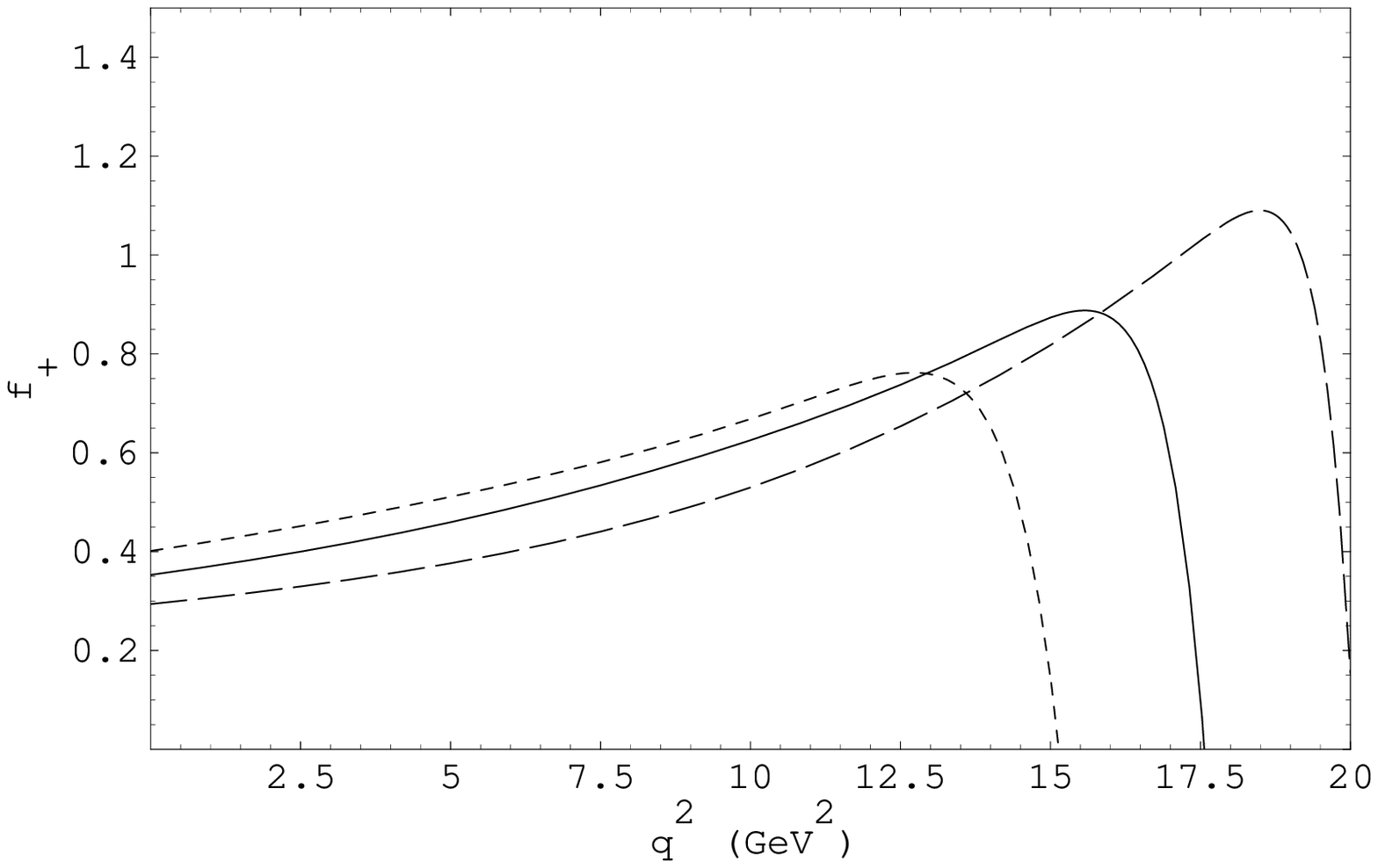}{Fig.5}

\PIC{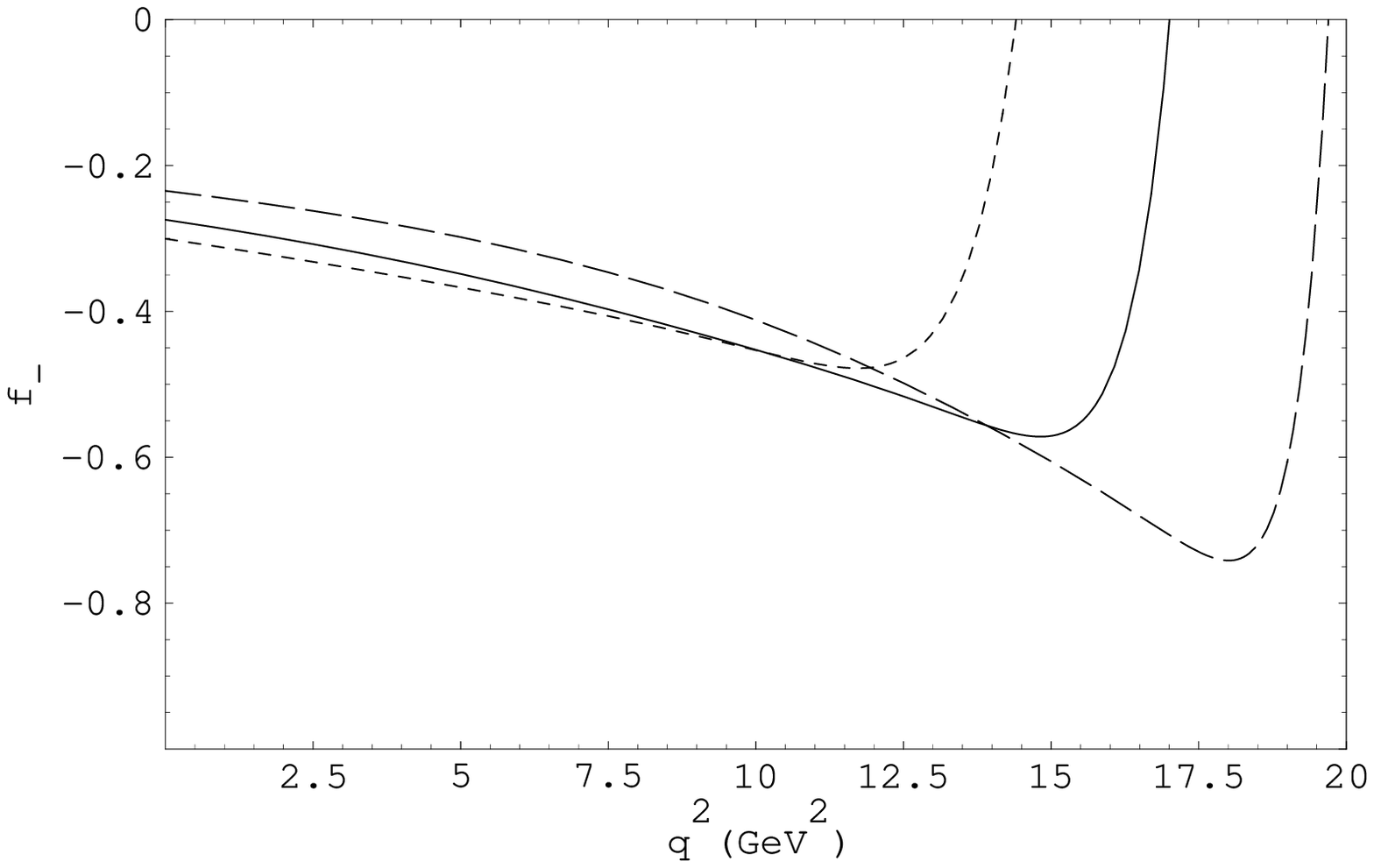}{Fig.6}

\vspace{-2cm}{\center{Fig.5-6. Variation of $f_+$ and $f_-$ with respect
to the momentum
transfer $q^2$ for different values of the continuum threshold $s_0$.
The dashed, solid and dotted curves correspond to $s_0=$1.7, 2.3 and 2.9 GeV
respectively. T=2.0 GeV is fixed.} }

\newpage
\mbox{}
\vspace{2cm}

\PIC{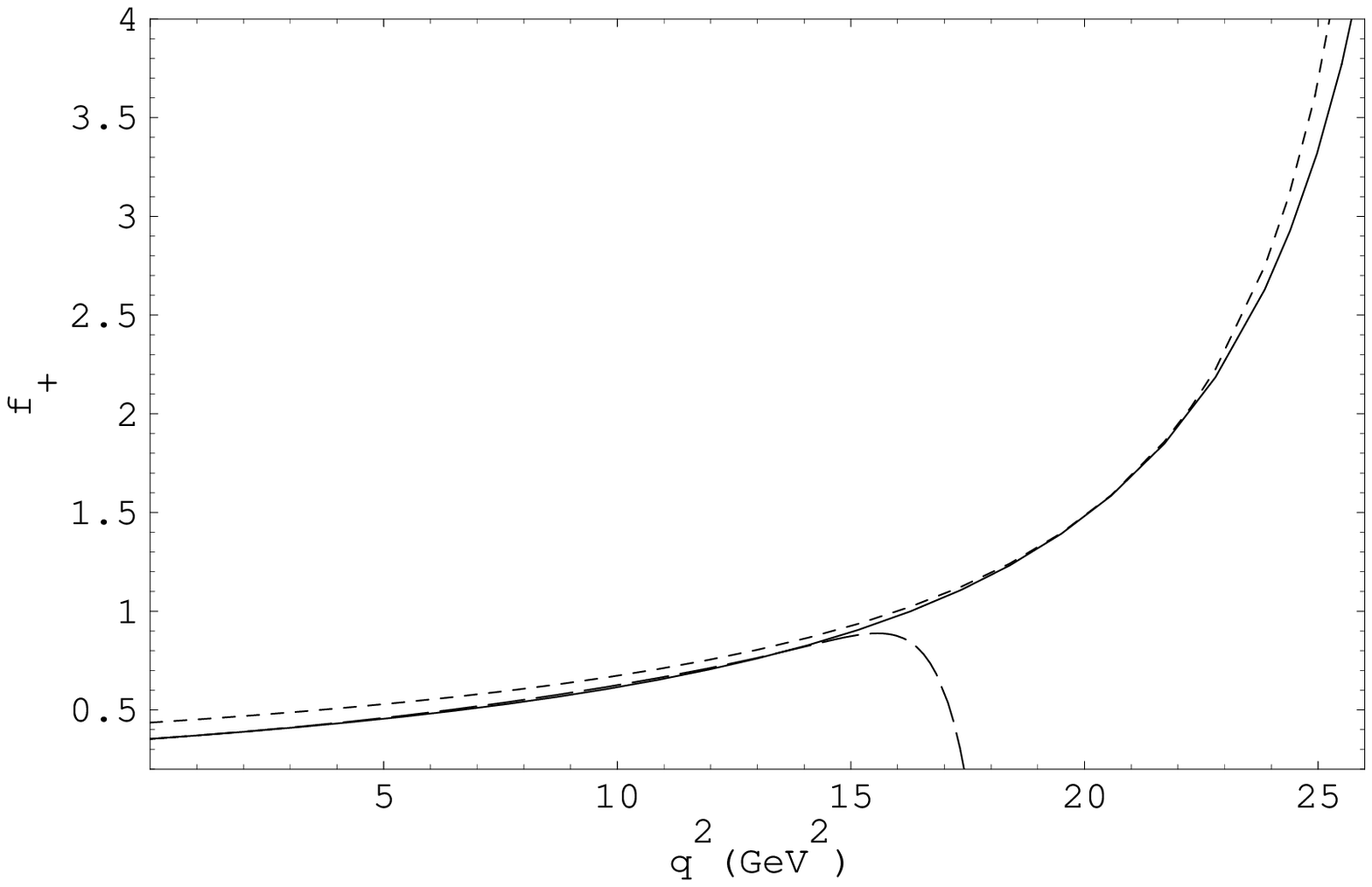}{Fig.7}

\vspace{-2cm}{\center{Fig.7. Variation of $f_+$ from different estimations.
The dashed curve is calculated from the sum rules
(\ref{sr}) for $s_0=2.3$GeV and $T=2.0$GeV. The dotted curve
comes from the single pole model (\ref{sinpole}).
And the solid curve is the result of (\ref{fitform}), which we used to evaluate
the integrated decay width and $|V_{ub}|$. The dashed line and the solid line
almost overlap each other at $q^2<15\mbox{GeV}^2$.} }

\end{document}